\documentclass[%
 aps,
 pre,%
 showpacs,
 amsmath,amssymb,
preprint,%
]{revtex4}
\usepackage{color}
\usepackage{graphicx}
\usepackage{dcolumn}
\usepackage{amssymb}
\usepackage{bm}

\newcommand{ \be}{\begin{equation}}
\newcommand{ \ee}{\end{equation}}
\newcommand{\beq}{\begin{eqnarray}}
\newcommand{\eeq}{\end{eqnarray}}
\newcommand{\bem}{\begin{pmatrix}}
\newcommand{\eem}{\end{pmatrix}}
\newcommand{\bmx}{\begin{array}}
\newcommand{\emx}{\end{array}}

\begin{document}

\title{Chromatic patchy particles: effects of specific interactions  on liquid structure}

\author{Oleg A. Vasilyev}
\affiliation{
 Max-Planck-Institut f{\"u}r Intelligente Systeme,
 Heisenbergstra{\ss}e~3, Stuttgart, Germany}
\affiliation{ IV. Institut f\"ur Theoretische Physik,
Universit{\"a}t Stuttgart,  Pfaffenwaldring 57, Stuttgart, Germany}

\author{Boris A. Klumov}
\affiliation{ Joint Institute for High Temperatures, Moscow,  Russia}
\affiliation{Institute for Information Transmission Problems,  Moscow, Russia}
\affiliation{Moscow Institute of Physics and Technology, Moscow, Russia}

\author{Alexei V. Tkachenko}
\affiliation{Center for Functional Nanomaterials, Brookhaven National Laboratory, Upton, NY,  USA}

\date{\today}

\begin{abstract}

We study the structural and thermodynamic properties of  patchy particle liquids, with a special focus on the role of "color", i.e. specific interactions between individual patches.
A possible  experimental realization of such "chromatic" interactions 
is by decorating the particle patches  with single-stranded DNA linkers.  The complementarity of the linkers can  promote selective bond  formation  between  predetermined pairs of patches. By using  MD simulations, we compare the local connectivity, the bond orientation order, and other structural properties of the aggregates formed  by the "colored" and "colorless" systems. The analysis is done for spherical particles with two different patch arrangements (tetrahedral and cubic). It is found that the aggregated (liquid) phase of the "colorless" patchy particles is better connected,  denser and typically has stronger local order than the corresponding "colored" one. This, in turn,  makes the colored liquid less stable thermodynamically. Specifically, we predict that in a typical case the chromatic interactions should increase the relative stability of the crystalline phase with respect to the  disordered liquid, thus expanding its region in the phase diagram.

\end{abstract}

\maketitle

\section{ Introduction}

In  recent years, systems of  patchy particles  have emerged among the key platforms  for advanced self-assembly \cite{patchy1}-\cite{VKT}. These are  typically  micron-scale colloids  featuring chemically distinct regions (patches) arranged in a pre-engineered pattern on the particle surface. In most  cases, the patches preferentially bind each other, thus giving rise  to  strongly anisotropic interparticle interactions, reminiscent of  covalent bonding in chemistry. Furthermore, by decorating the  patches with single-stranded DNA molecules, one can introduce multiple types of patches as well as   a selective type-dependent binding through DNA hybridization\cite{Patchy DNA}. This can be interpreted as  "coloring" of patchy particles. One can expect that adding  "color" to the directionality  of the  interactions would  lead to a greater control over the resulting morphology.

While the current interest in  patchy colloidal  systems is motivated primarily by their  potential for  programmable self-assembly of ordered structures, the study of their disordered phases is of great conceptual importance as well  \cite{Sciort_2013}-\cite{VKT}. In particular, this  provides valuable insights both into the equilibrium phase behavior and  kinetics of self-assembly. For instance, Smallenburg and Sciortino   \cite{Sciort_2013} have recently demonstrated that the ground state of a system  of patchy particles need not  be a crystal, even when the particles themselves are highly symmetric. Specifically, for the case of  four-patch particles with tetrahedral symmetry, the Cubic Diamond (CD) crystal becomes the thermodynamically preferred state only in the limit of   strong  bond directionality, i.e. a very small patch size. Otherwise, the system can achieve its maximum connectivity without sacrificing all of its configurational  entropy, and thus preserving the  liquid-like order.  A similar conclusion was reached independently in our recent study \cite{VKT}. Rather than exploring the equilibrium phase behavior, we were interested in the structural properties of the random liquid-like aggregate formed from a low density gas of patchy particles. Remarkably, the four-patch system demonstrates a relatively  high degree of local ordering upon aggregation, as opposed to the six-patch particles  with  cubic symmetry. We  explained this difference by observing that  the the coordination number $Z=4$  is the maximum connectivity of  a  disordered aggregate under an  assumption that the patches are point-like but  the bond directionality is not extreme. Therefore, the four-patch particles may have all their patches connected without forming a crystalline state, in contrast to the  six-patch system.  

In this work, we  expand the previous analysis to  study the  effects of  "chromatic" interactions between the patches on the structural and thermodynamic properties of the liquid phase. One might  expect that the bond directionality when combined  with the color-based selectivity
(e.g., due to DNA functionalization of the patches) should  lead to  stronger  ordering and better programmability of the self-assembled structures \cite{AT1}-\cite{AT2}. As in our previous study, we analyze both  4-patch particles with  tetrahedral symmetry (4pch)  and   6-patch particles with  cubic symmetry (6pch). For each of  these systems, we  consider two extreme cases: when  all the patches are equivalent ("colorless" system), and when  all patches that belong to  the same particle are of  different colors, and their pairwise  interactions are subject to a complementarity rule ("colored" system).

\section{ Model and numerical algorithm}

In our model, we describe a patchy particle
as a solid sphere of diameter $\sigma$. Patches are located on the surface
of this sphere and rotate with it.

The motion of a single particle is represented as a combination of a translational displacement of its center
and a rotation of the particle around it. In order to  take into account the rotational degrees of freedom, we describe the orientations of the particles with quaternions.

\begin{figure*}
\includegraphics[width=0.4\textwidth]{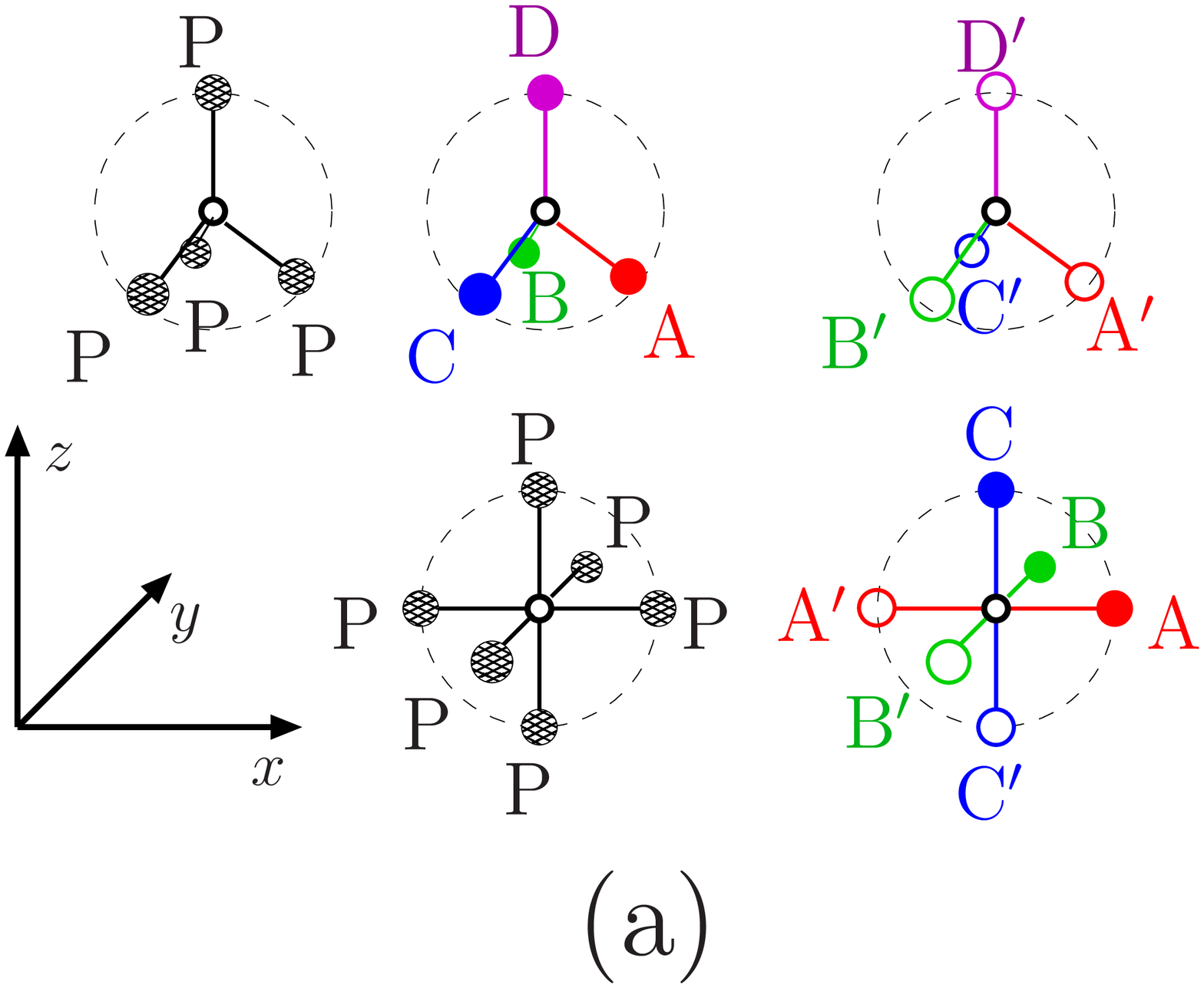}
\includegraphics[width=0.4\textwidth]{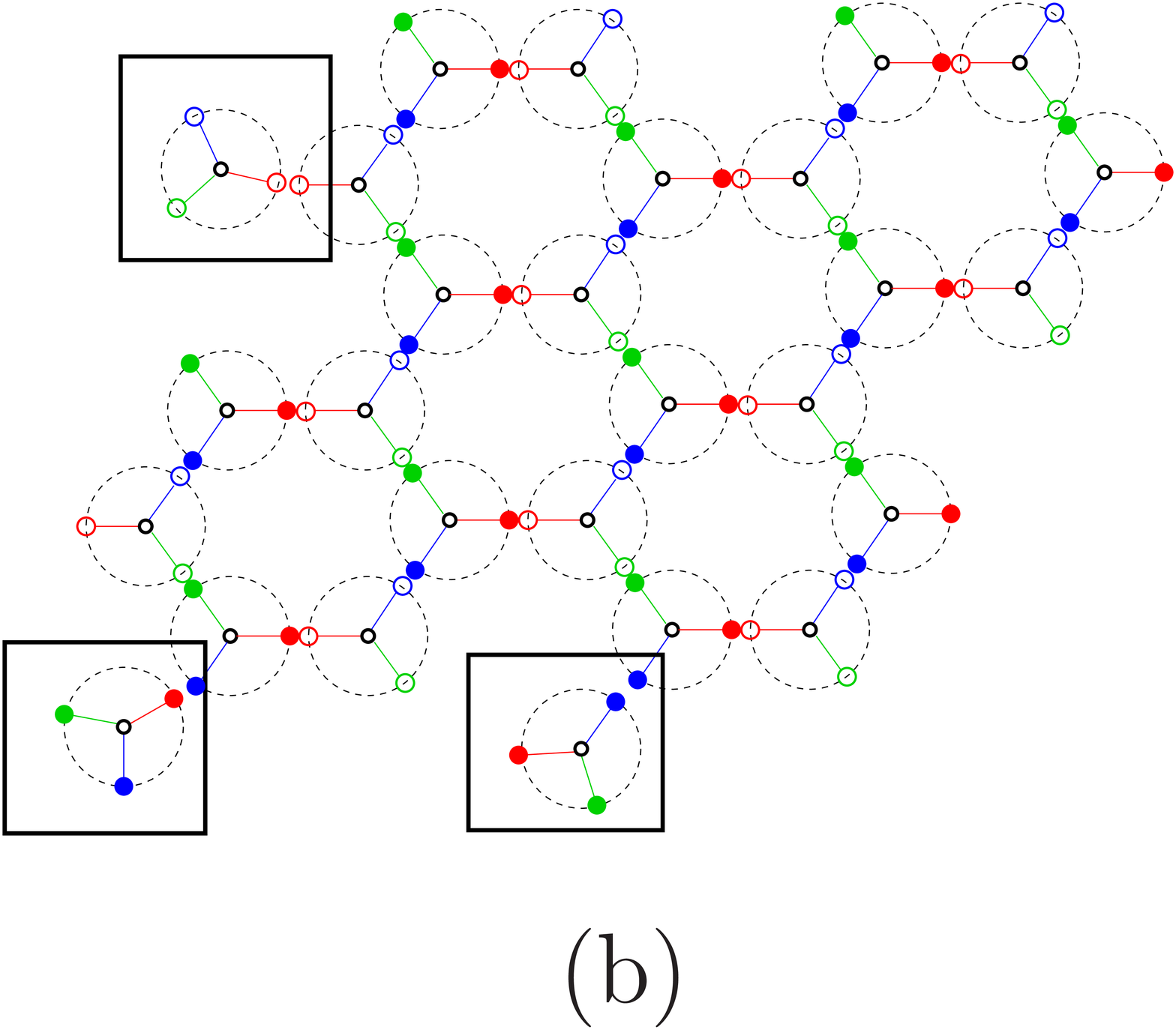}
\includegraphics[width=0.43\textwidth]{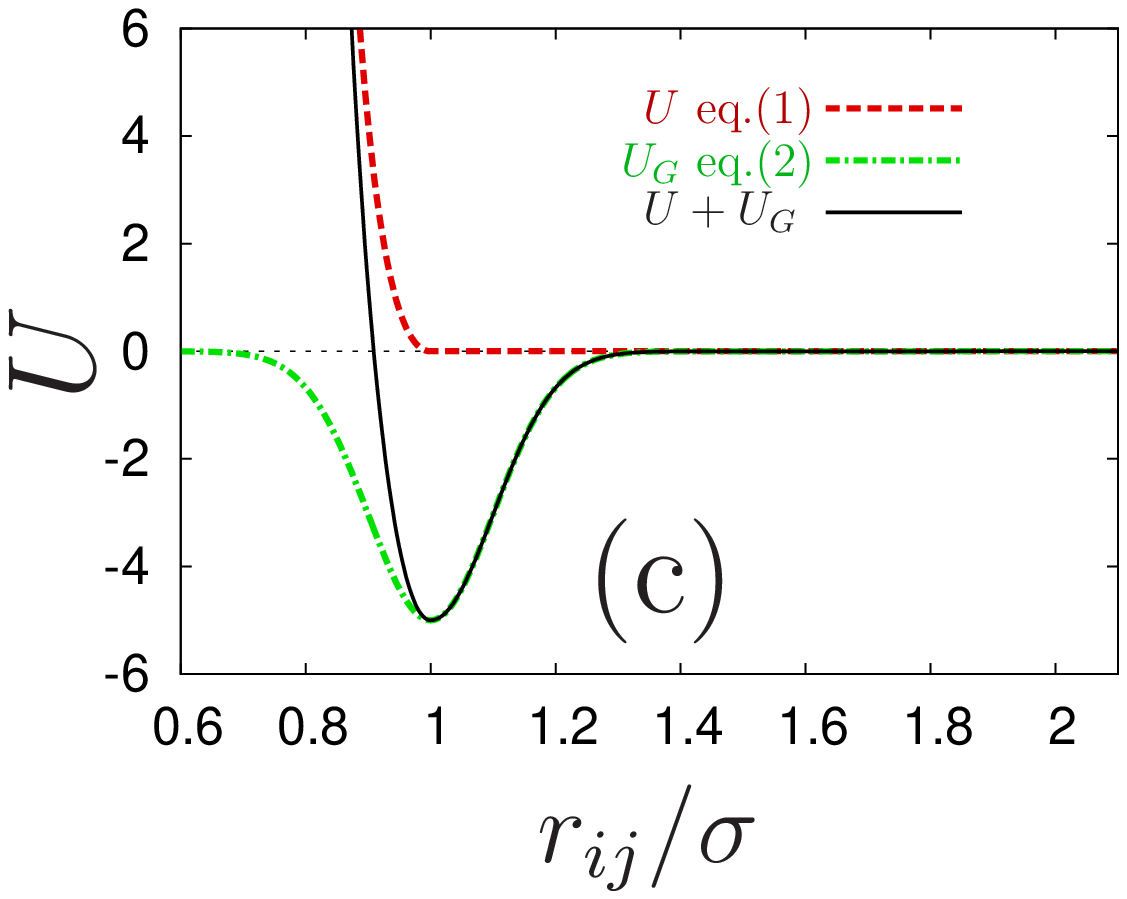}
\includegraphics[width=0.43\textwidth]{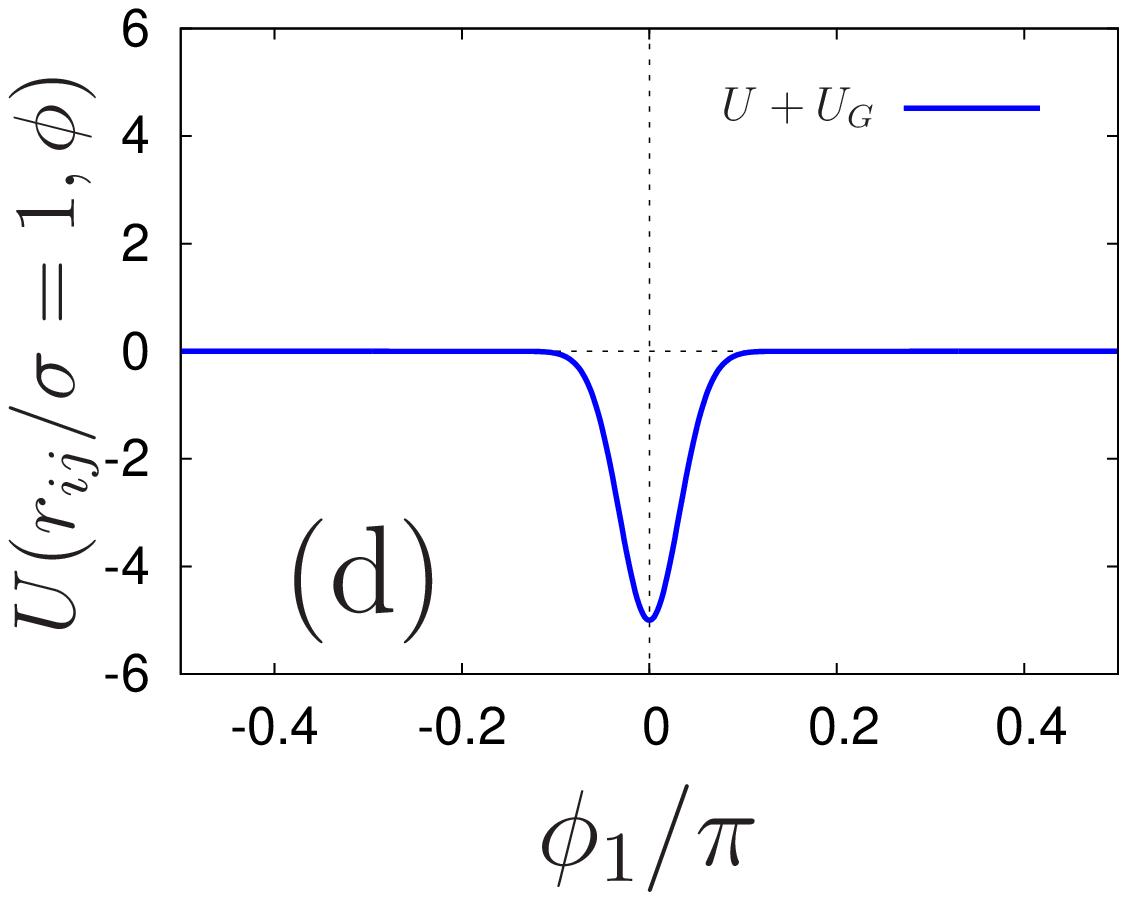}
\caption{(Color online) (a) Arrangement of the colorless and  colored patches for 4pch and 6pch particles;
(b) Top view of a  layer of a perfect CD crystal for the colored 4pch  system.
Particles in black squares do not interact with the aggregate
due to the complementarity rule of chromatic interaction;
(c) Interaction potentials $U$ (Lennard-Jones core-core repulsion Eq.~(\ref{eq:Ulj})), $U_{G}$ (Gaussian patch-patch attraction Eq.~(\ref{eq:UG})) and the total potential $U+U_{G}$ plotted  as  functions of the normalized 
center-to-center distance $r_{ij}/\sigma$,  for interaction strength $U_{p}=5$;
(d) The angular dependence of  interaction potential $U$ on  normalized orientation angle $\phi{1}/\pi$ of the first particle with respect to the center-center line. The interparticle distance is fixed at $r_{ij}/\sigma=1$, $U_{p}=5$.}
\label{fig:geom}
\end{figure*}

Interactions between the  patches satisfy the following rules.  
In  the colored case, only complementary pairs of  patches
$AA'$, $BB'$, $CC'$, and $DD'$ attract each other.
Interactions between other,  non-complementary,  pairs such as 
$AA$, $AB'$, etc., are absent.
In the case of colorless particles,  every pair of patches $PP$
are mutually attractive.

At the start of simulation,  the position of the $k$-th patch
of the $j$-th particle with respect to the particle  center is given
by  vector ${\bf a}_{j}^{(k)}(0)$. The system of colored 4pch particles is binary.
A "primary" 4pch particle has the following patch positions:
${\bf a}^{(1)}_{j}(0)= \left( \sqrt{\frac{2}{9}}\sigma,0,-\frac{1}{6}\sigma\right)$ (A type),
${\bf a}^{(2)}_{j}(0)= \left( -\sqrt{\frac{1}{3}}\sigma,\sqrt{\frac{1}{3}}\sigma,-\frac{1}{6}\sigma\right)$ (B type),
${\bf a}^{(3)}_{j}(0)= \left( -\sqrt{\frac{1}{3}}\sigma,-\sqrt{\frac{1}{3}}\sigma,-\frac{1}{6}\sigma\right)$ (C type), and
${\bf a}^{(4)}_{j}(0)= \left( 0,0,\sigma/2\right)$ (D type).
 A "complementary" particle  is a mirror image of a "primary" one, with  patches $A',B',C',D'$ located at the positions
 $A,C,B,D$, respectively (see Fig.~\ref{fig:geom}(a)). For  the colored 6pch particle,  the  complementary patches are located opposite to each other:
${\bf a}^{(1,2)}_{j}(0)= \left( \pm \sigma/2,0,0\right)$ for $A$ and $A'$,
${\bf a}^{(3,4)}_{j}(0)= \left(0, \pm \sigma/2,0\right)$ for $B$ and $B'$,
${\bf a}^{(5,6)}_{j}(0)= \left(0,0, \pm \sigma/2 \right)$ for $C$ and $C'$.

In our description, vector ${\bf r}_{j}(t)$  represents the position  of the particle $j$ at time $t$. The orientation of that particle is described by unit quaternion ${\bf \Lambda}_{j}(t)$.
This quaternion has a form ${\bf \Lambda}_{j}(t)=[\cos(\phi_{j}/2),\sin(\phi_{j}/2) {\bf n }_{j}(t)]$,
where  unit  vector ${\bf n }_{j}(t)$ defines the rotation axis  passing through the center of the particle and
$\phi_{j}$ is the angle of rotation around this axis (see,~e.g., Ref.~\onlinecite{EQ}).

The location of the $k$-th patch of that particle  is therefore given by
${\bf a}_{j}^{(k)}(t)={\bf r}_{j}(t)+{\bf \Lambda}_{j}(t) \otimes {\bf a}^{(k)}_{j}(0)
\otimes \tilde {\bf \Lambda}_{j}(t)$.
Here $\otimes$ denotes the quaternion product, and   $\tilde {\bf \Lambda}_{j}(t)=[\cos(\phi_{j}/2),-\sin(\phi_{j}/2) {\bf n }_{j}(t)]$  is the corresponding   conjugated quaternion.

This quaternion-based approach can be used   for Molecular Dynamic (MD) simulations of rigid objects~\cite{EM}.
 There are  two types of interactions in our model. Patchy particles repel each other with the following isotropic short-range potential:
\begin{equation}
\label{eq:Ulj}
U({\bf r}_{ij})=\left\{
\begin{array}{ll}
U_{0}({\bf r}_{ij})+(\sigma-|{\bf r}_{ij}|) U'_{0}(\sigma),& {\bf r}_{ij} \le \sigma\\
0,& {\bf r}_{ij}>\sigma\\
\end{array},
\right.
\end{equation}
where
$U_{0}({\bf r}_{ij})=4 \epsilon_{0} \left[
\left( \sigma/{\bf r}_{ij}\right)^{12}-
\left(\sigma/{\bf r}_{ij}\right)^{6}
 \right]$
 is the standard Lennard-Jones potential,
 $U'_{0}(\sigma)=\left. \frac{{\mathrm d}U_{0}(r)}{{\mathrm d} r}\right|_{r=\sigma}$
 is its derivative,
$\sigma$ is the interaction distance (equivalent to the particle diameter) as well as cut-off distance,
$\epsilon_{0}=1$ is the interaction  strength, and ${\bf r}_{ij}={\bf r}_{i}-{\bf r}_{j}$
is the vector connecting the centers
of the $j$-th and the $i$-th particles (see Fig.~\ref{fig:geom}(c), dashed line). The potential is expressed in units of $k_{\mathrm B}T$, which is the fundamental energy scale in our problem.  
In addition to this  isotropic interparticle repulsion, the patches
that belong to different particles and satisfy complementarity rule (in the colored case) attract each other with a  Gaussian potential:
\begin{equation}
\label{eq:UG}
U_{G}({\bf a}_{ij}^{(kl)})=-U_{p}
\exp \left[-\left({\bf a}_{ij}^{(kl)}\right)^{2}/2w^{2}\right]
\end{equation}
where ${\bf a}_{ij}^{(kl)}={\bf a}_{i}^{(k)}-{\bf a}_{j}^{(l)}$
is the vector connecting the patch $l$ of the particle $j$
and the patch $k$ of the particle $i$, $w=0.2$ is the half-width of the interaction
and $U_{p}$  is the strength of the interaction in units of $k_{\mathrm B}T$.
In  Fig.~\ref{fig:geom}(c) we plot both potentials (for  $U_{p}=5$) as well as their sum
$U+U_{G}$ (shown by the solid line). The potentials are shown as 
 functions of the normalized distance between the two
particle centers $r_{ij}/\sigma$  for the case when the centers and the attracting patches are located on the same  line. The cut-off distance for the Gaussian potential is $5w$.
In Fig.~\ref{fig:geom}(d)  the angular dependence of  potential $U$ is plotted 
for a fixed  interparticle distance, $r_{ij}/\sigma=1$
(Fig.~\ref{fig:geom}(c) corresponds to $\phi_{1}/\pi=0$).

From a known set of displacements and orientations of all particles
$\{{\bf r}_{j},{\bf \Lambda }_{j} \}$
one can  compute a set of total forces
and torques $\{{\bf F}_{j},{\bf M }_{j} \}$ acting on  them:
\be
\label{eq:n}
\left\{
\begin{array}{l}
 \dot {\bf v}_{j} (t) =\frac{ 1}{m}
 {\bf  F}_{j}(\{ {\bf  r}_{j},{\bf \Lambda}_{j}\})\\
\dot { \bf \omega }_{j}(t)=\frac{ 1}{I}
 {\bf  M}_{j}(\{ {\bf r}_{j},{\bf \Lambda}_{j}\}) \rule{0pt}{12pt}\\
\end{array}
\right.,
\left\{
\begin{array}{l}
  \dot { \bf r }_{j}(t)  =
{\bf v}_{j}(t)\\
\dot {\bf  \Lambda }_{j}( t)=\frac{1}{2}
{\bf  \omega}_{j}(t) \otimes  {\bf \Lambda }_{j}(t)
\end{array}
\right.
\ee
Here $I$ is the moment of inertia,
$\dot {\bf v}_{j}$ and $\dot { \bf \omega }_{j}$ are linear
and angular accelerations, respectively.
All lengths are measured in  units of  particle radius  $\sigma/2$.
The mass of a particle is $m=1$, the moment of inertia of a solid sphere is
$I=\frac{1}{10}m \sigma^{2}=0.4$.

We use the Verlet 
numerical algorithm with the time step $dt=0.002$ for the  numerical integration of Eq.~(\ref{eq:n}). 
The coupling to a thermal bath  is represented by the Langevin noise added to the forces and torques in Eq.~(\ref{eq:n}):\ ${\bf  F}_{j}=-\gamma {\bf v}_{j}(t)+\xi_{j}(t)$
and ${\bf  M}_{j}= -\frac{1}{3}\gamma \sigma^{2} {\bf \omega}_{j}(t)+\zeta_{j}(t)$. Here $\gamma=3 \pi \nu \sigma$ is the friction coefficient for the solvent viscosity $\nu$. The strengths of delta-correlated  noise terms $ \xi_{j}(t)$ and $ \zeta_{j}(t)$ are set by the  Fluctuation-Dissipative theorem:  $\left<  \xi_{i}^{\alpha}(t)\xi_{j}^{\beta}(t') \right>=
2\gamma \delta_{i,j}\delta_{\alpha,\beta}\delta_{t,t'} $,
$\left<  \zeta_{i}^{\alpha}(t)\zeta_{j}^{\beta}(t') \right>=
\frac{2}{3}\gamma \sigma^{2}  \delta_{i,j}\delta_{\alpha,\beta}\delta_{t,t'} $.
In our units ($\sigma/2=1$, $m=1$, $k_{\mathrm B}T=1$),  the friction coefficient $\gamma$
has a  meaning of the diffusion time of a particle over the distance equal to  its radius.  
On the other hand,  the time constant $\tau=1/\gamma $
corresponds to the crossover from the  ballistic to the diffusion regime of  motion. We set  $\gamma=10$, so that   the dynamics of a particle is Brownian on times
$t \gg \tau=0.1$.

In our simulations,  the control parameter is 
 the interaction strength   $U_{p}$.
 For small values of $U_{p}$, the system is in a gas phase;
 for large values of $U_{p}$, it exhibits an aggregation into an amorphous "liquid" phase.
We simulate a set of $N=1000$ particles in a cubic
box of size $L=48$ with periodic boundary conditions. The volume fraction for this system is $\eta= \pi \sigma^{3} N/(6 L^{3})\simeq 0.038$. To verify the robustness of our results, we have  performed additional simulations of the "colored" 4pch system for volume fractions  twice as large, $\eta \simeq 0.072$,  and half the reference value, $\eta \simeq 0.019$. This did not result in any significant variation, aside from the logarithmic shift of the value of   $U_p$  at which  aggregation is observed, in a good agreement with  theoretical expectations.  Indeed, in a simple example of the reaction ``monomer+monomer $\leftrightarrow$ dimer'',  the density renormalization  $\eta \to A \eta $ can be offset by a logarithmic shift of  the interaction potential  $U_p \to U_p- k_{\mathrm B}T\ln(A)$. A similar logarithmic shift is expected for the case of coexistence of an extended aggregate with a gas of monomers.  The simulation time is set to  t=10000, which is sufficient for most observables to reach their saturation values. Note that the inter-patch binding remains reversible at this time scale, except for very high  interaction  strength,   $U_p>15$.

\begin{figure}
\includegraphics[width=8.4cm]{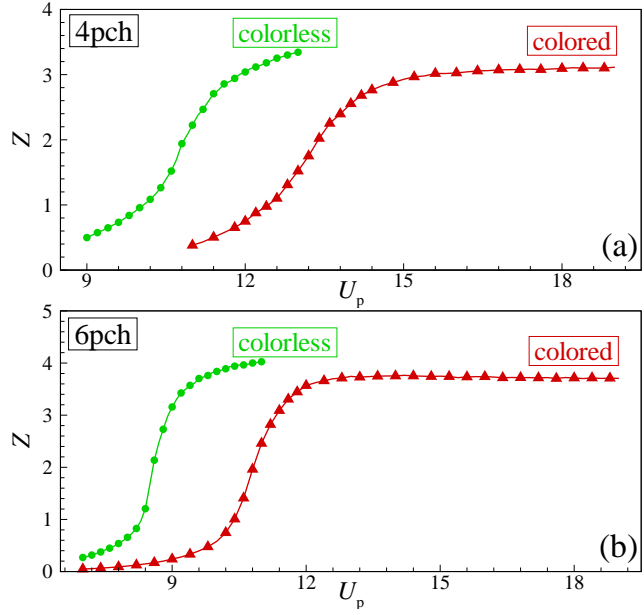}
\caption{(Color online) Average  number of the  topological nearest neighbors $Z$ versus   interaction potential $U_p$ for the colored (red triangles) and the colorless (green circles) systems. The data for 4pch  and 6pch systems are presented in plots (a) and (b), respectively.}
\label{cl6}
\end{figure}

\begin{figure}
\includegraphics[width=8.4cm]{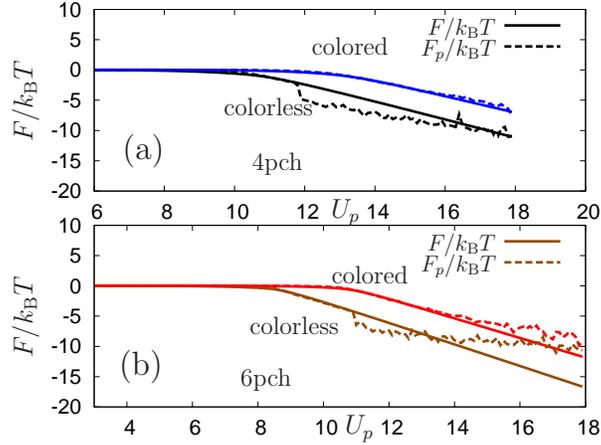}\\
\caption{(Color online) Helmholtz free energy per particle $F$ calculated by the thermodynamic integration and $F_{p}=1+\ln(p)-p$ (in $k_{\mathrm B}T$ units) plotted  as  functions of  $U_{p}$, for the colored and the colorless systems of (a) 4pch and (b) 6pch particles.}
\label{integration}
\end{figure}
 
\begin{figure}
\includegraphics[width=8.4cm]{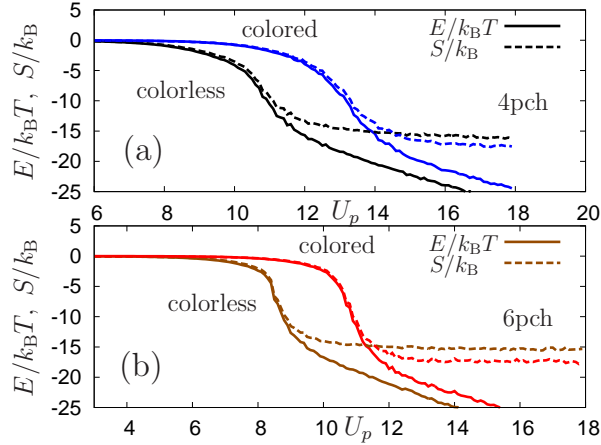}\\
\caption{
(Color online) Comparison of the energy per particle $E$ (in $k_{\mathrm B}T$ units) and the entropy per particle $S$ (in $k_{\mathrm B}$ units) plotted as functions of  $U_{p}$ for the  colored and the colorless systems of (a) 4pch and (b) 6pch particles.}
\label{ES}
\end{figure}

\begin{figure}

\includegraphics[width=8.4cm]{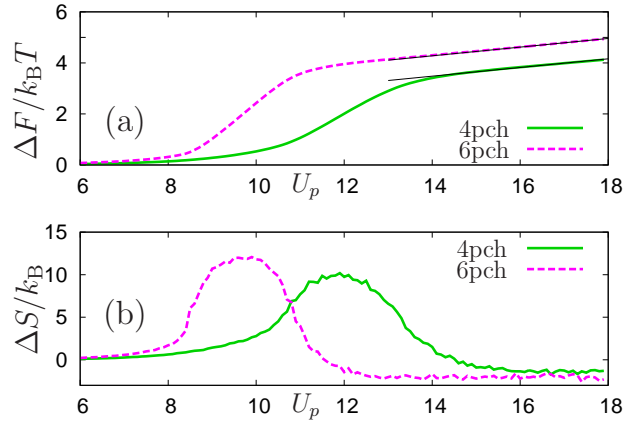}\\
\caption{(Color online) (a) The difference $\Delta F$ (in $k_{\mathrm B}T$ units) between the Helmholtz free energy per particle of the  colored and  colorless systems vs. $U_{p}$,  for 4pch and 6pch particles.
(b) The difference $\Delta S$ (in $k_{\mathrm B}$ units) between the entropy per particle of the colored and  colorless systems vs. $U_{p}$, for 4pch and 6pch particles.
}
\label{DES}
\end{figure}

\begin{figure}
\includegraphics[width=8.4cm]{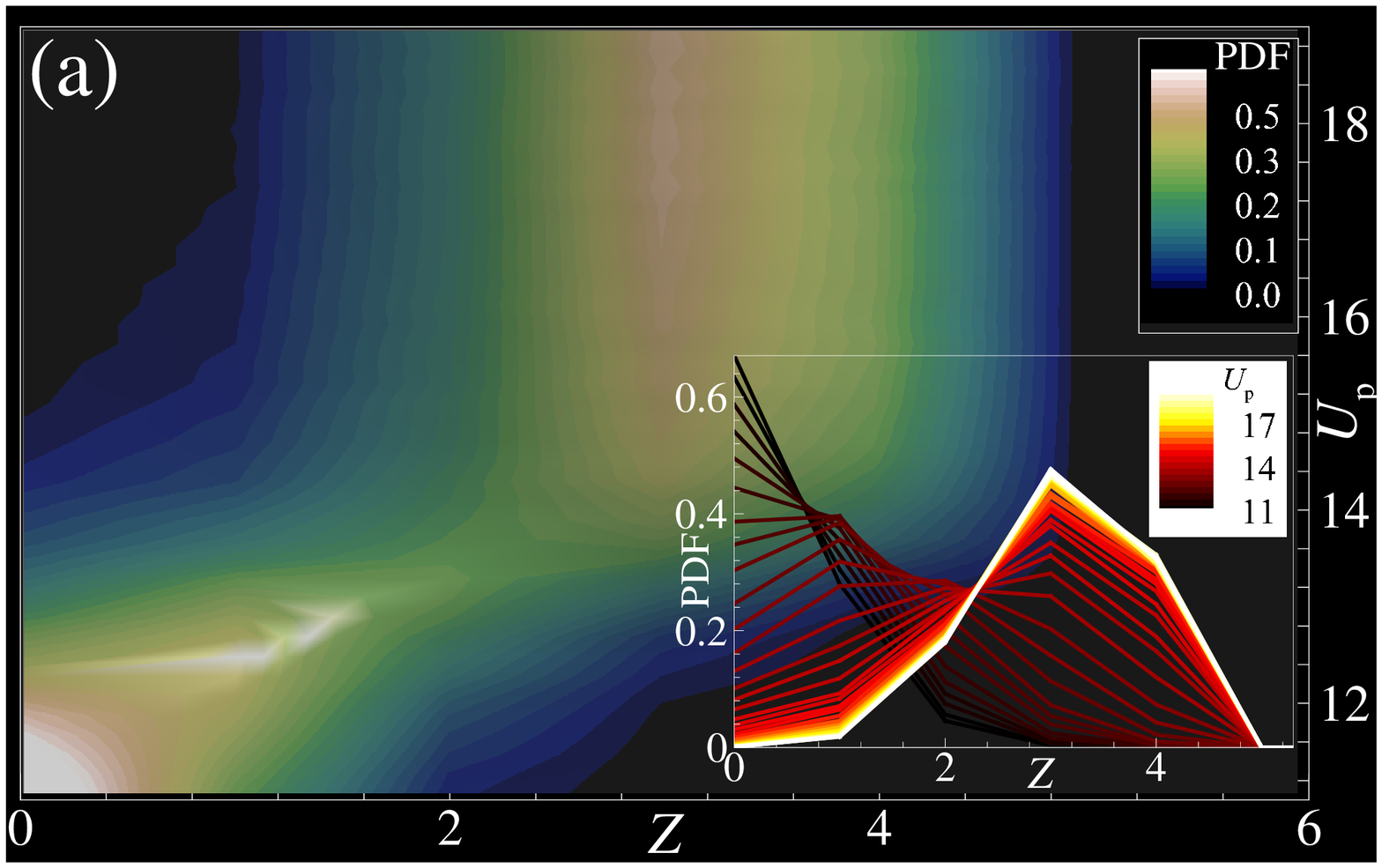}\\
\includegraphics[width=8.4cm]{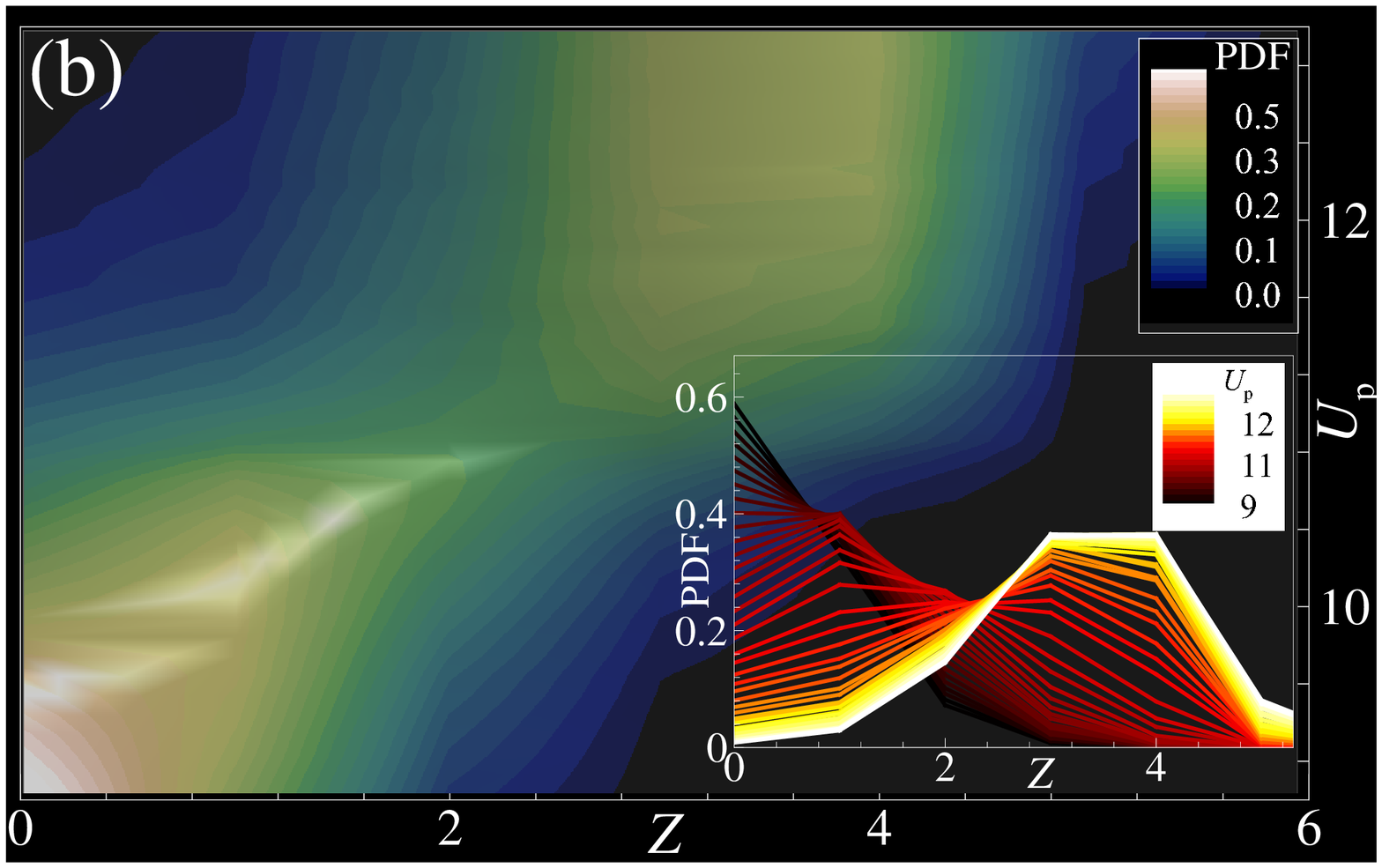}\\
\caption{(Color online) The probability distribution  function (PDF) of the coordination number $Z$  at variable  interaction strengths $U_p$ (vertical axis). Plots (a) and (b) represent the colored and colorless 4pch systems, respectively.  The insets show  the same PDFs at selected values of $Up$ (the bright color corresponds to the strongest attraction).} \label{cl1}
\end{figure}

\
\begin{figure}
\includegraphics[width=8.4cm]{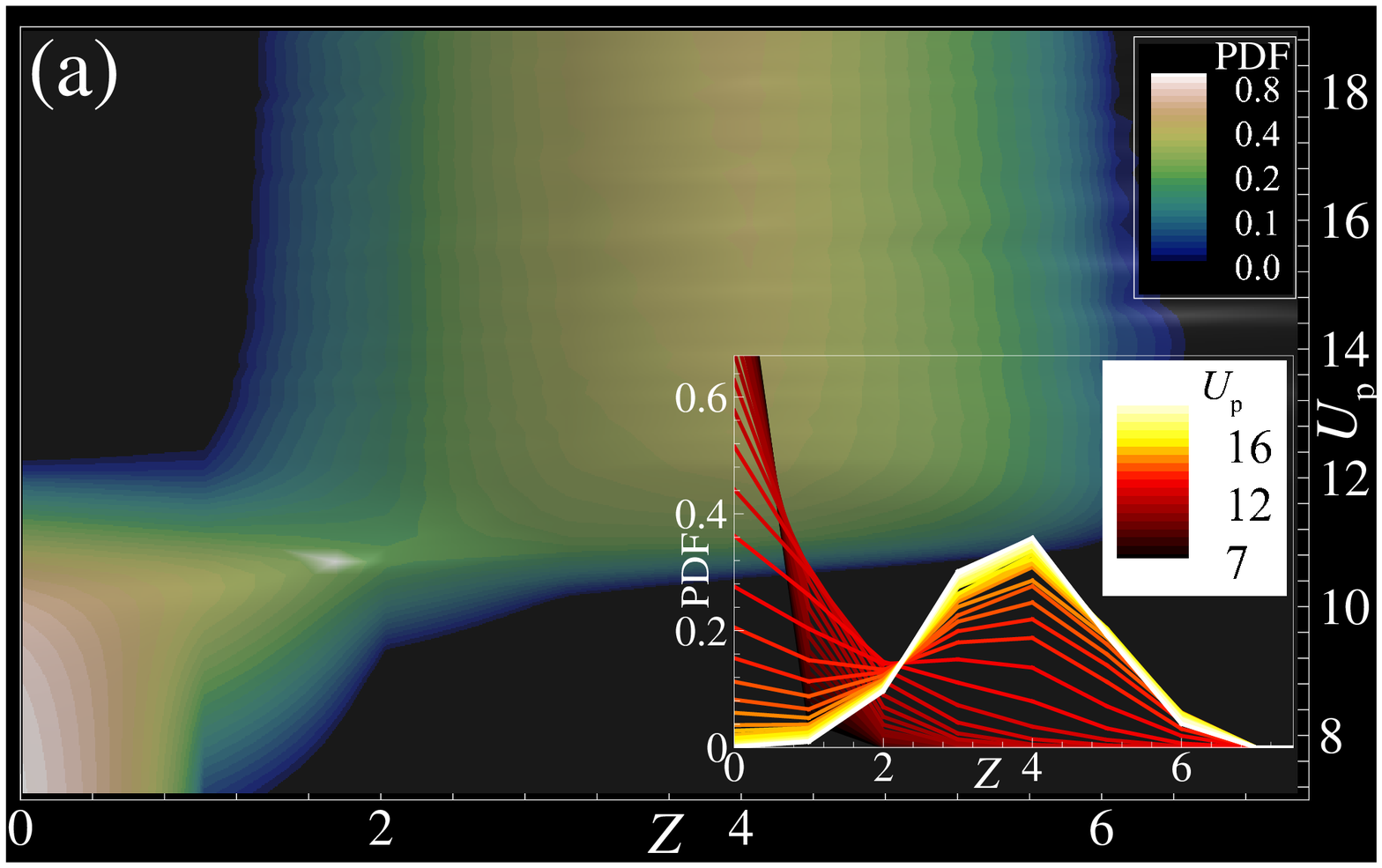}\\
\includegraphics[width=8.4cm]{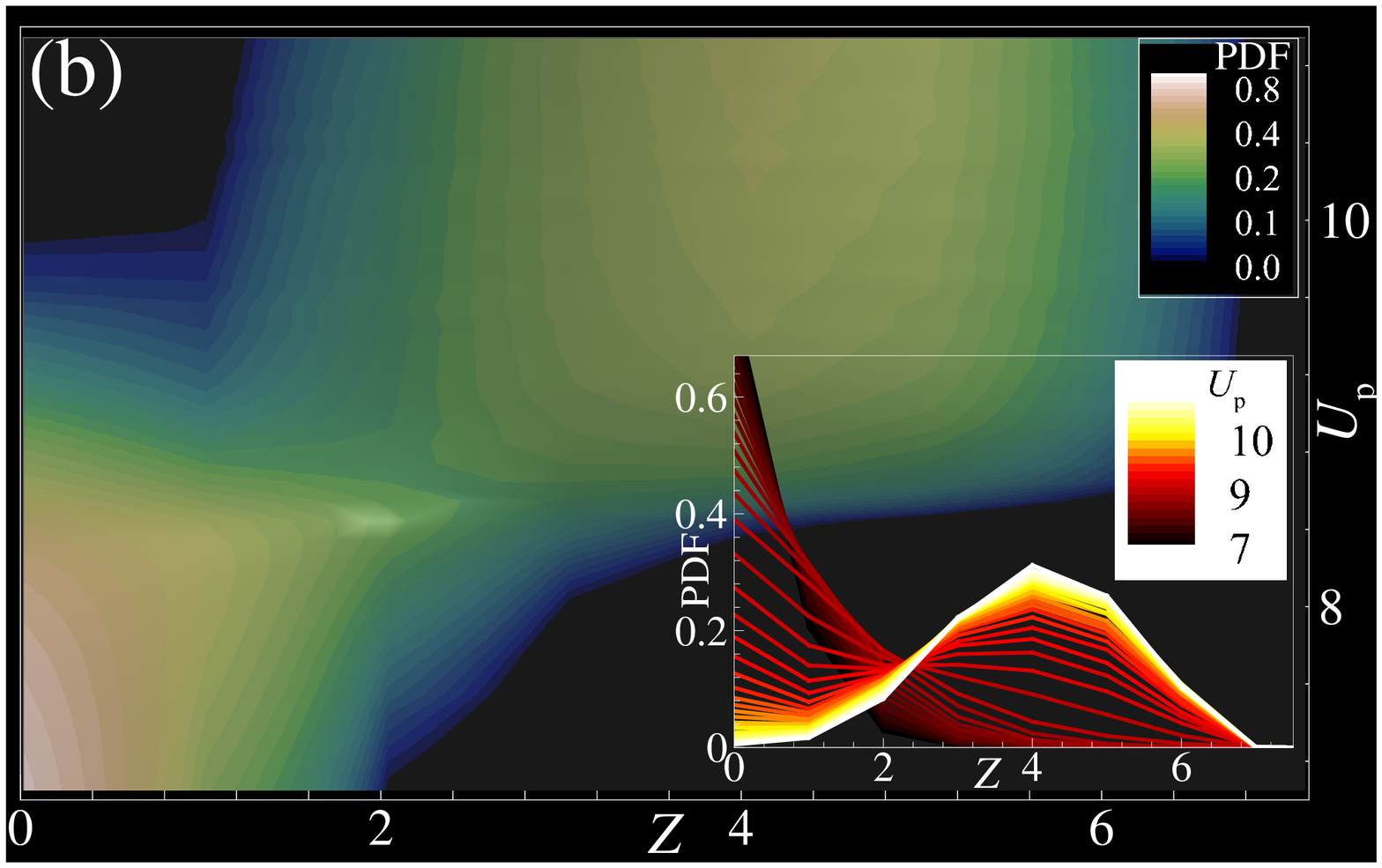}\\
\caption{(Color online) The probability distribution  function (PDF) of the coordination number $Z$  at variable  interaction strengths $U_p$ (vertical axis). Plots (a) and (b) represent the colored and colorless 6pch systems, respectively.  The insets show  the same PDFs at selected values of $Up$ (the bright color corresponds to the strongest attraction).}
\label{cl3}
\end{figure}

\section{Results and Analysis}
\subsection{Topological Characterization}

Two patches are considered to be bound if the distance 
between their centers is less than the half-width $w=0.2$.
We define a pair of  particles to be topological nearest neighbors (NNs) if there is such a bond  formed between their  patches. This definition is robust with respect to the choice of the threshold length $w$ since the interacting patches are strongly localized relative to each other. The computed  values of the  mean coordination number $Z$ (i.e. the number of NNs) are  plotted in Fig.~\ref{cl6} vs.  interaction strength $U_{p}$, 
for 4pch (a) and 6pch (b) systems. As  $U_{p}$ increases, all the systems undergo  transitions from a gas to an amorphous aggregate.

From the  thermodynamic point of view, the instability of the uniform gas of particles towards the formation of a dense aggregate corresponds to its spinodal decomposition. When this occurs, the aggregate is essentially an early liquid phase. The reversibility of the interparticle bonding (at $U_p<15$) implies  that the  local structure of the aggregate is similar to the bulk liquid. It should be noted, however, that   the relaxation of clusters  to  spherical droplet shapes is a much slower process not captured in our study. In view of this, and due to the finite system size,  there is always an interfacial correction to any of the thermodynamic observables reported below.  Partially because of this limitation, we do not use our data to directly compare the thermodynamics  of the liquid and crystalline phases. Instead, we are focused on a specific  aspect of the problem: the effect  of chromatic interactions on  structural and thermodynamic properties of the liquid.

The aggregation  is reflected by a gradual change of the  coordination number from  $0$ (unbound particles), to a certain saturation value $Z_0$. The latter is the characteristic of the liquid phase and its value depends on the system. In particular, both   colored and colorless 6pch particles form aggregates with $Z_0$ close to 4 (4.2 and 3.8, respectively).  4pch systems exhibit a somewhat lower connectivity: $Z_0 \simeq 3.2$ for the colored and $Z_0 \simeq 3.5$ for the colorless one.  

The intermediate values of the mean NN number correspond to the coexistence of the condensed liquid phase with a gas of single particles and smaller clusters.  All these cases support our earlier argument that the coordination number $Z_0=d+1$ plays a special role for a patchy liquid in d-dimensional space \cite{VKT}. This value emerges  from the comparison of the number of mechanical constraints with the number of translational degrees of freedom. The rotational degrees of freedom are not taken into account in this counting since a much stronger patch-patch localization is required in order for them to be frozen. The fact that $Z_0$ is somewhat  below 4 for 4pch system is also natural as $Z=4$ is the maximum possible number of bonds for these particles, and  some of the patches remain unbound, especially for the colored system. One could expect  $Z=4$ to be reached in the limit of very strong attraction, but that regime is not accessible partially because of the finite size  effects, and also since the binding becomes nearly irreversible on the timescale of the simulations when the interactions are stronger than  the threshold value, $U_p\simeq 15$.       

\subsection{Chromatic Effect on Thermodynamics}

The colored patchy  particles  exhibit  aggregation at a  significantly higher interaction strength $U_{p}$ than  the corresponding colorless ones. Co-existence of the liquid-like aggregate with the gas of particles implies that  the chemical potential  is the same in  both phases. Therefore, onset of the condensation approximately corresponds to the point at which the  chemical potential of the free particle gas,  $\mu_0=k_{\mathrm B}T \log \eta$ is  equal to that of the liquid phase.  The  shift of the aggregation curves, $\Delta U_p$,    between the colorless and the corresponding colored system   shown in   Fig.~\ref{cl6},      can be  related to the difference in  the  chemical potential $\Delta \mu$ between the two  aggregates, taken at the same   interaction potential.  Indeed, $\Delta \mu$ should be  offset by the difference in the interaction energy per bond:
\be
\frac{\Delta \mu}{k_{\mathrm B}T}=\frac{Z_0}{2} \Delta U_p\;.
\label{eqil}
\ee
 Here,   $\frac{Z_0}{2}$ is the average number of bonds per particle in the aggregate.

  This estimate gives  $\Delta \mu \simeq 4 k_{\mathrm B}T$  for both 4pch and 6pch systems.   More reliably the  free energy of the aggregated  phase can be found by the  thermodynamic integration~\cite{TI}:
  \be
F=\int_0^{1/k_{\mathrm B}T}\frac{E(\beta k_{\mathrm B}TU_p)d \beta}{\beta}  =\int_0^{1}\frac{E(x U_p) d x}{x} \;.
\label{integr}
\ee
Here, $E(U_p)$ is the ensemble averaged potential energy per particle, and $F$ is the Helmholtz free energy per particle. The uniform  gas phase  that corresponds to $U_p=0$  is chosen as  a reference state,  $F=0$.  In this calculation we took into account the fact  that $U_p$ is  expressed in units of  $k_{\mathrm B}T$. We ignore the kinetic energy contribution to $F$ since it is completely decoupled from any structural transformation in non-quantum systems. 
 
The chemical potential $\mu$ can be calculated  as the Gibbs free energy per particle. In the limit of a completely aggregated state (high value of $U_p$), the pressure coming from the dispersed particles can be neglected, and the Helmholtz and Gibbs free energies  become identical; i.e.  $F=\mu$. According to the results of thermodynamic integration,   presented in Fig. ~\ref{integration}, the chemical potential of the colored aggregate is increased  with respect to the colorless one  by an amount well described by a linear function of $U_p$:
   \be
\Delta \mu=\Delta F= \Delta_0+0.17 k_{\mathrm B}TU_p
\label{Delta}
\ee
 Here, $\Delta_0=1.1k_{\mathrm B}T$ for 4pch and $\Delta_0=1.9k_{\mathrm B}T$ for 6pch.  In both these cases, this  "chromatic" correction is close to $4k_{\mathrm B}T$  near  the aggregation   point, in  a  good  agreement with our earlier rough estimate. 
 Furthermore, the result of integration can be independently related  to the fraction of particles in the gas phase, $p$.  Namely, the chemical potential of a gas of unbound particle can be expressed in terms of  their volume fraction, $p \eta$ as  $\mu=k_{\mathrm B}T\log(p \eta)$. The Helmholtz free energy per particle differs from the  chemical potential by the  amount $PV/N$, where  pressure $P$ is proportional to the number density of all the dispersed objects, including the  individual particles and the bigger clusters.  In the regime when this correction is relevant (i.e. of the order of $k_BT$), the pressure is  dominated by the contribution from the gas of  single particles,  $P=pNk_{\mathrm B}T/V$.  Therefore,  $\mu-F=PV/N=pk_{\mathrm B}T$.  This leads to an  alternative expression for $F$:
  \be
F_p=k_{\mathrm B}T(\log(p)-p+1)
\label{Fp}
\ee
Here, we have subtracted the Helmholtz   free energy of the uniform gas phase, to keep the same reference state as in Eq. (\ref{integr}).
While this method is significantly  less accurate than the thermodynamic integration (as can be seen  in Fig.~\ref{integration}), it  provides an additional verification for  our results.  

One can also  separate the  entropic and energetic  contributions to the  chromatic free energy correction,  Eq. (\ref{Delta}),  by noting that $TS=E-F$.  The corresponding  data are presented in Fig.~\ref{ES}. For both 4pch and 6pch systems, the entropic correction $TS$ reaches a constant value very close to  the corresponding parameter $\Delta_0$ in Eq.~(\ref{Delta}). Hence,  the linear term represents the energy correction, due to the lower number of NNs in the colored case.  Within this interpretation, the  coefficient $0.17$ should be close to $\Delta Z_0/2$,  implying that  $\Delta Z_0 \simeq 0.35 $ is the  difference in the mean coordination number between the respective colored and colorless systems. The value of this parameter is indeed consistent with the  data in Fig.~\ref{cl6}. 
The free energy difference $\Delta F$ (in $k_{\mathrm B}T$ units) and the entropy difference 
$\Delta S$ (in $k_{\mathrm B}$ units)
for the colored and colorless systems are plotted in Fig.~\ref{DES}.

\subsection{Comparison to Crystal Phase}
 
The computed effect of the chromatic interactions  on the thermodynamics of the  liquid phase can be compared to the similar effect expected in the  crystal phase. Note that the  colored and colorless systems  when arranged  into an ideal crystalline configuration are locally indistinguishable: they have the same ground state energy and the same vibrational modes, unless the coloring is inconsistent with the given crystal lattice. Therefore, the only difference in their free energies is due to  configurational entropy. Specifically, that the colorless particles have a larger number of equivalent orientational states that would preserve the bonds with their neighbors.  There are 12 such
 orientations of a colorless 4pch particle  and 24 for the case of   6pch system. The resulting  difference in the chemical potential between the  colorless and colored system is purely entropic,  $k_{\mathrm B} T\log 12 \simeq 2.5k_{\mathrm B}T$ and   $ k_{\mathrm B}T\log 24 \simeq 3.2k_{\mathrm B}T$,   respectively. 
 
 On  the one hand, this entropic correction is  greater than that  computed  for the corresponding  liquid phases ($\Delta_0=1.1 k_{\mathrm B}T$ and $\Delta_0=1.8 k_{\mathrm B}T$). However, this entropic effect is easily offset by the energy difference. Indeed, according to Eq.~(\ref{Delta}), $\Delta \mu$ becomes greater that the corresponding  correction for the crystal ($2.5k_{\mathrm B}T$ and $3.5k_{\mathrm B}T$ respectively) once the interaction parameter exceeds  $U_p\simeq 8$. While the parameter itself is model specific, this interaction strength is definitely insufficient to condense the dispersed particle system even at the relatively high  volume fraction used in our simulations. 
 
 We therefore expect that in a generic case where the patch-patch interaction is sufficient  for the aggregation of a dilute  colloidal suspension, the  ordered phase will gain additional stability due to the chromatic interactions.   
 In other words, if we start with the colorless system, and make the patches colored with the color-specific binding, this would lead to  free energy penalty both to the crystal and liquid phases, but the latter penalty would be  greater. The colored and colorless systems are certainly physically distinct, so these free energy differences are not directly  measurable. However, we use the gas phase at the same concentration as a reference state for all the cases, which allows us to determine the effect of coloring on  the relative stability of the liquid and crystal phases.

 It should be emphasized that we do not perform a direct comparison of the liquid to the crystal, instead we only evaluate  the strengths of the chromatic effects for each of the phases separately. While the one for the  crystal is obtained based on an analytic argument, the effect on the bulk liquid is extracted  from our MD simulations and therefore may have errors coming from interfacial effects. One can  estimate the role of the interface by noting that the  average coordination number of colorless 4pch aggregate, $Z_0=3.5$ deviates from its maximum possible value 4 by approximately $10\%$. This sets an upper bound on the strength the interfacial/finite size  correction  to  the overall free energy when compared to the bulk liquid. We therefore expect that the presence of an extended interface should result  a similar correction to the chromatic effect itself. Such a correction would not alter our main conclusions in any significant manner. 

\subsection{Local Connectivity and Correlations} 

Fig.~\ref{cl6} indicates only  a modest difference in the maximum
values of the mean coordination number  $Z$ between the colored and
colorless systems.  However, the detailed analysis of its PDFs reveals a much  greater qualitative contrast. 
Figs.~\ref{cl1}-\ref{cl3} show the  influence of the color of the patches on the distribution of the number of  NNs   for   4pch and 6pch systems, respectively. As  interaction  strength  $U_p$ increases, the systems form amorphous  aggregates  \cite{VKT} with  nearly constant distributions of the topological NNs. This transition is clearly seen in Figs.~\ref{cl1}-\ref{cl3}  for all the four cases.

The colored 4pch system is  dominated by particles with $Z=3$,
while its colorless counterpart has a much larger fraction of
particles with completely saturated bonds, i.e. $Z=4$.
This must be a result of additional  constraints imposed by
the coloring of the patches. A similar trend, although not as pronounced, is also seen for the 6pch particles.

\begin{figure}
\includegraphics[width=8.4cm]{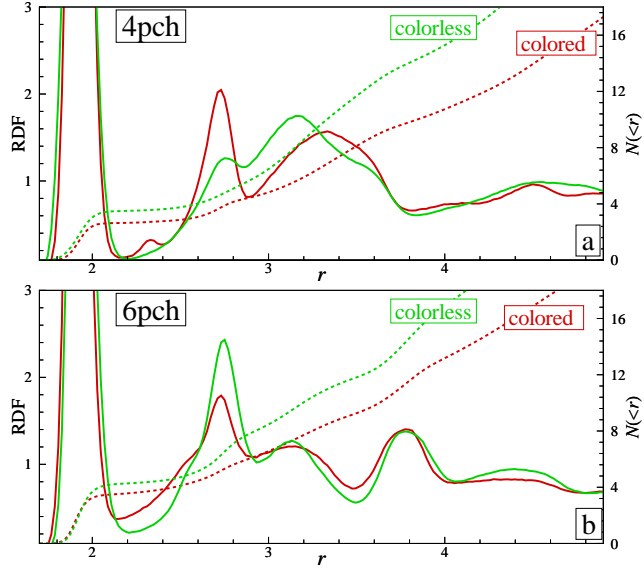}
\caption{(Color online) The radial distribution function (RDF) $g(r)$ and the corresponding cumulative function $N(<r)$ ($N(<r) \equiv 3 \eta \int_0^r \xi^2 g(\xi) d\xi$), showing the  mean number of particles inside the sphere of  radius $r$ plotted for colored (red curves) and colorless (green curves) liquid-like systems 4pch (a) and 6pch (b). It is clearly seen that the  colorless patchy systems form denser aggregates than the colored ones.}
\label{cl8}
\end{figure}

An alternative characterization of the local structure in the aggregated liquid phase  is provided  by  the radial distribution function $g(r)$,  shown in Fig.~\ref{cl8} for the 4pch and 6pch systems, both colored (red curves)
and  colorless (green curves). Additionally, the average  number of  particles inside of a sphere of radius $r$,  $N(<r) \equiv 3 \eta \int_0^r \xi^2 g(\xi) d\xi$, are  presented on the same plots.  There is a clear correspondence between the topological NNs discussed above, and the geometrical NNs represented by the first peak in $g(r)$: their number is also close to $Z=4$, and it is slightly lower for the colored system than for the colorless one.
The analysis of the cumulative curves  reveals that  aggregates of the  colorless patchy particles  are somewhat  denser than those of colored ones, which is consistent with our observation that they are also more connected.

\subsection{Orientational Order}

To further characterize the local structural properties of the patchy system, we use the bond orientational order
parameter method \cite{stein}, which has been widely used in the context of condensed matter physics \cite{stein}, hard sphere~\cite{hs1,hs2,hs3} and Lennard-Jones systems~\cite{lja,ljb,ljn,ljc}, complex plasmas~\cite{cp0,cp1,cp2,pu}, colloidal suspensions \cite{coll1,coll2}, metallic glasses~\cite{mg}, confined films~\cite{nc,ncc}, granular media, etc. Within this method, the rotational invariants  of rank $l$ of both the second $q_l(i)$ and the third $w_l(i)$ order are calculated for each particle $i$ in the system from the vectors (bonds) connecting its center with the centers of its $N_{\rm nn}(i)$ nearest neighboring particles
\be
q_l(i) = \left ( {4 \pi \over (2l+1)} \sum_{m=-l}^{m=l} \vert~q_{lm}(i)\vert^{2}\right )^{1/2}
\ee
\be
w_l(i) = \hspace{-0.8cm} \sum\limits_{\bmx {cc} _{m_1,m_2,m_3} \\_{ m_1+m_2+m_3=0} \emx} \hspace{-0.8cm} \left [ \bmx {ccc} l&l&l \\
m_1&m_2&m_3 \emx \right] q_{lm_1}(i) q_{lm_2}(i) q_{lm_3}(i),
\label{wig}
\ee
\noindent
where $q_{lm}(i) = N_{\rm nn}(i)^{-1} \sum_{j=1}^{N_{\rm nn}(i)} Y_{lm}({\bf r}_{ij} )$, $Y_{lm}$ are the spherical harmonics and
${\bf r}_{ij} = {\bf r}_i - {\bf r}_j$ are the vectors connecting the centers of particles $j$ and $i$. We note that the bond order parameters $w_l$ scale as  $w_l \propto q_l^3 $; so, in general, these parameters are much more sensitive to the local orientational order in comparison with $q_l$. Here, to define structural properties of  patchy particles,  we calculate the rotational invariants $q_l$, $w_l$ for each particle that has the maximum number of NNs:  $Z = 4, 6$, respectively. The first shells of the ideal 4pch and 6pch  systems correspond to the  cubic diamond (CD) and the simple cubic (SC) arrangements, respectively. 

The values of the corresponding rotational invariants $q_l$ and $w_l$ for the perfect patchy crystals  are shown in Table 1. 
\begin{table}[!ht]
  \centering
  \caption{Rotational invariants for perfect patchy crystals}\label{t:t1}
 \begin{tabular}{ll|cccc}
\hline\hline
system & structure & \quad $q_{4}$ & \quad $q_{6}$ & \quad $w_{4}$ & \quad $ w_{6}$ \\ \hline
4pch & CD (1st shell, 4 NNs) & 0.509 & 0.628 & -0.159 & -0.013 \\
6pch & SC (1st shell, 6 NNs) & 0.76 & 0.35 & 0.159 & 0.013 \\

\hline\hline
\end{tabular}
\end{table}

\begin{figure}
\includegraphics[width=8.4cm]{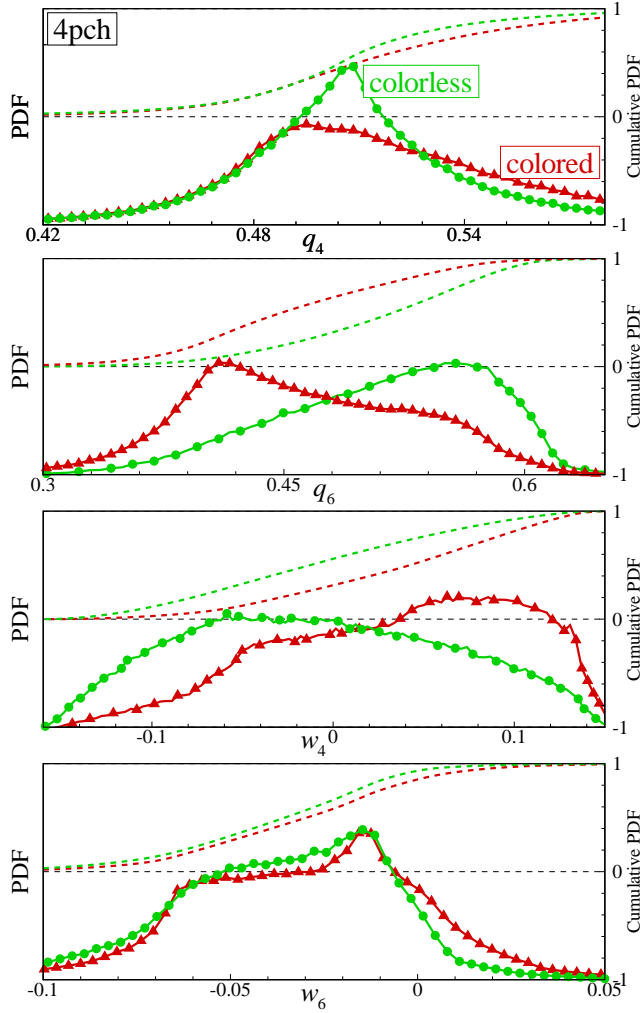}
\caption{
(Color online)
Patchy system 4pch. Probability distribution functions $P(q_l)$ and $P(w_l)$ of   different rotational invariants $q_l $ and $w_l$ ($l=4,~6$),  calculated for  particles with exactly  $Z =4$ topological NNs.  PDFs are  plotted  both for the  colored (red triangles) and  colorless (green circles) patchy systems in the aggregated  phase.}
\label{cl9}
\end{figure}

\begin{figure}
\includegraphics[width=8.4cm]{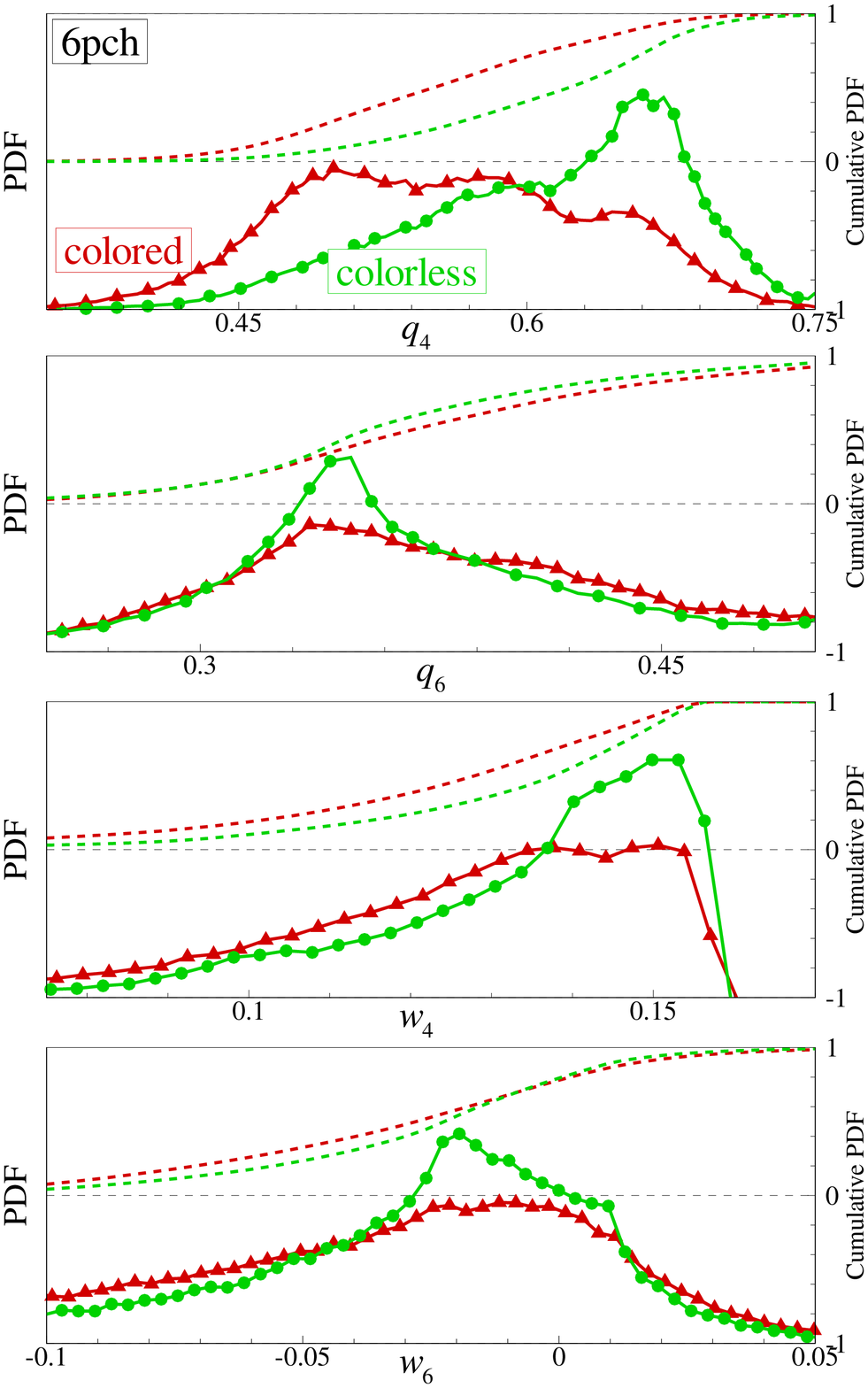}
\caption{
(Color online) Patchy system 6pch. Probability distribution functions $P(q_l)$ and $P(w_l)$ of  different rotational invariants $q_l $ and $w_l$ ($l=4,~6$),  calculated for particles with exactly  $Z =6$ topological NNs.  PDFs are  plotted  both for the colored (red triangles) and colorless (green circles) patchy systems  in the aggregated  phase. }
\label{cl10}
\end{figure}

The characterization  of the local bond  orientation order for 4pch and 6pch aggregates is presented in Fig.~\ref{cl9} and Fig.~\ref{cl10}  respectively, both for the  colored (triangles) and the colorless (circles) cases. The probability distribution functions $P(q_l)$ and $P(w_l)$ of the  rotational invariants $q_l $ and $w_l$ ($l=4,~6$), and their cumulative distributions shown in the figures, reveal that the colorless systems are  more ordered than the colored ones.  Specifically, for the  colorless system the PDFs exhibit broad yet pronounced  peaks at the locations expected for the corresponding perfect crystal. The colored liquid is significantly less ordered, which is in a general agreement with the observation that it is less connected and more frustrated due to the color constraints, and with the fact that it is less stable thermodynamically. 

\begin{figure}
\includegraphics[width=8.4cm]{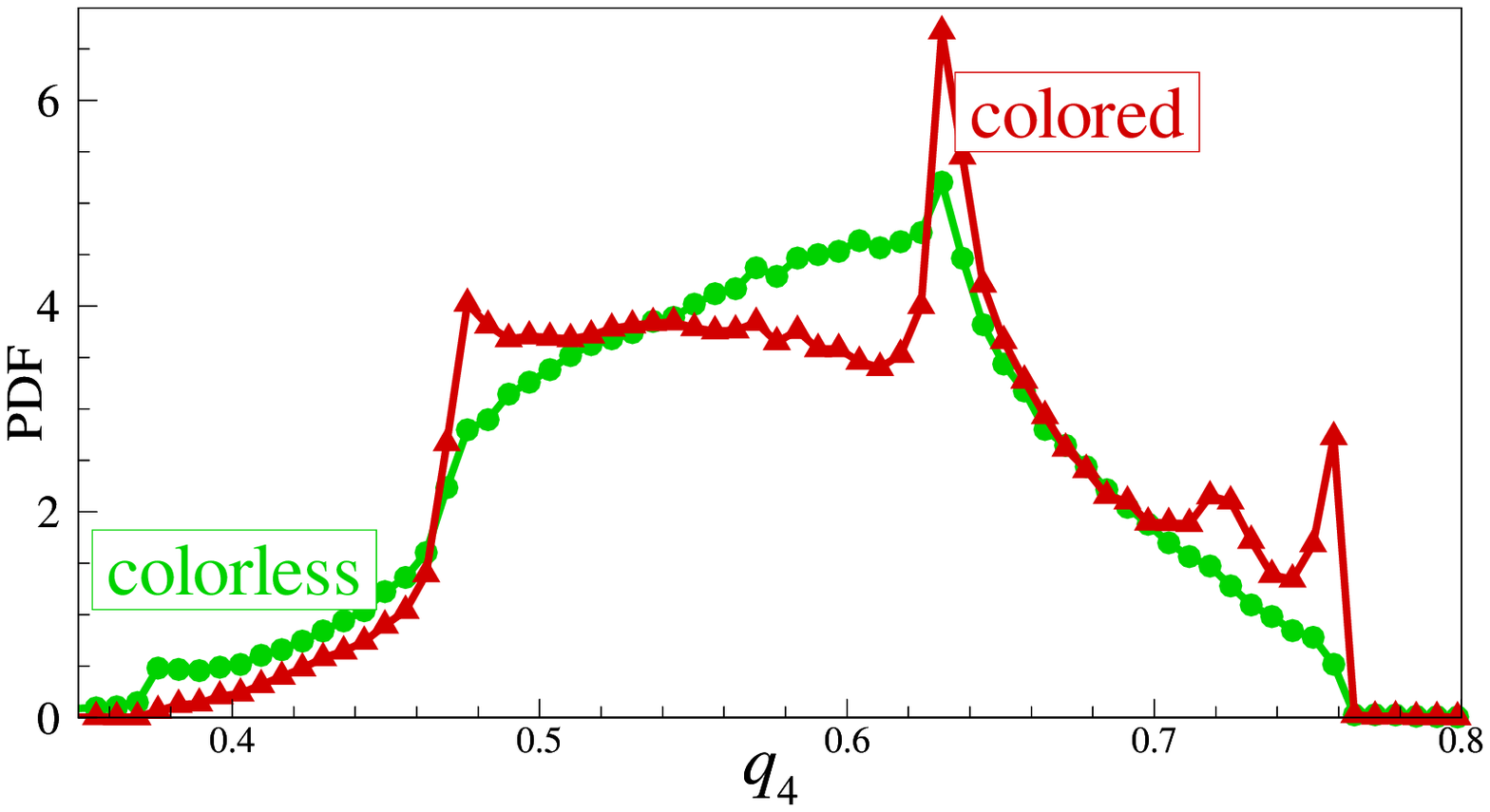}
\caption{(Color online) Patchy system 4pch. Probability distribution function $P(q_4)$  calculated for particles with exactly  $Z =3$ topological NNs, both for the  colored (red triangles) and colorless (green circles)  systems  in the aggregated  phase. Note the strong peaks indicating better local ordering of the colored system, in striking contrast with the results obtained from the analysis of fully connected particles with $Z=4$, in Fig \ref{cl9}.}
\label{cl11}
\end{figure}

This mostly  coherent picture is however challenged when we analyze the local bond orientation order for the subset of 4pch particles with a smaller  number of NNs, $Z=3$. 
In particular,  we distinguish a striking example of $P(q_4)$ that indicates remarkable  enhancement of the local order in the colored system
compared to the colorless one (see Fig.~\ref{cl11}). A possible explanation of this anomaly is that 
additional bonds  in the colorless system come at the
expense of  deformation of the network, and the resulting strain affects weaker connected $Z=3$ nodes  more strongly  than those with $Z=4$. Indeed, if we imposed a
condition that each interparticle bond is perfectly aligned with
the vector pointing towards the center of the patch, the maximum
average coordination number that could be achieved would be  $Z=5/3$.
As we have discussed in our earlier work ~\cite{VKT},
this constraint is typically unrealistic, and the coordination
number is expected to be larger (tending towards $Z=4$,
consistently with the present results). These additional bonds
however do result in a modest bond rotation and strain in the particle network.
Presumably, this leads to the  suppression
of local order near $Z=3$ particles in the better connected colorless system.

\section{ Conclusions}

The central conclusion of our study is that the colored patches provide an important additional element of control over the self-assembled structures. In our work, we considered  the  early stages of  liquid formation from a gas of patchy particles, both colored and colorless. We analyzed the structural and thermodynamic properties of the aggregated phase and  concluded that the colorless patchy systems are typically  more connected, denser and more ordered than the corresponding colored ones. The only identified  anomaly that goes against this general trend deals with the local bond  orientation order near 4pch particles with  coordination number $Z=3$. That type of order is noticeably increased when the  chromatic interactions are introduced. 

By means of  thermodynamic integration, we were able to compute the chromatic correction to the chemical potential of the  aggregated phase, which we treat as a model for the bulk liquid. On the other hand, we evaluated  a similar correction to the crystal phase based on an analytic entropic argument. By comparing the two results, we predict that in a generic case,  the chromatic interaction between  patches would lead to a greater free energy penalty for the disordered liquid phase than for the corresponding ordered structure.    
This implies an enhanced stability of  crystals made of the colored patchy particles compared to the uncolored ones.  In particular, we expect that the  limitation on the patch size required for the crystallization according to 
Ref.~\onlinecite{Sciort_2013}  would be  less severe in the chromatic case.

The chromatic interactions discussed in this work may be employed not only to shift the balance in the  order-disorder transitions, but also to select a desired  self-assembled  structure in the case of crystalline polymorphism. The classical example of that kind  is the competition between cubic and hexagonal diamond lattices (CD and HD), both of which have identical tetrahedral  arrangement  of NNs. It is an attractive idea to select a specific lattice by chromatic patch-patch interactions. Unfortunately, our binary system of 4pch particles  does not provide  sufficient control for that: both CD and HD are possible with the same set of particles.   However, it should be possible to control the outcome in  polymorphic system if more than two particle types are used. This constitutes an intriguing conceptual problem for  future study.

This study is supported by European Research Council under FP7 IRSES
Marie-Curie grants PIRSES-GA-2010-269139 and PIRSES-GA-2010-269181.
Research carried out in part at the Center for Functional
Nanomaterials, Brookhaven National Laboratory, which is supported by the U.S.
Department of Energy, Office of Basic Energy Sciences, under Contract No. DE-AC02-98CH10886.
BAK was  supported by  the Russian Science Foundation, Project no. 14-12-01185.

\end{document}